\documentclass[preprint,12pt]{elsarticle}
\usepackage{amssymb}
\usepackage{amsmath}
\usepackage{float}
\usepackage{graphicx}
\usepackage{subcaption}
\journal{Nuclear Physics B}

\begin{document}

\begin{frontmatter}



\title{Charm production and fragmentation fractions at midrapidity in pp collisions at the LHC with ALICE}


\author{Renu Bala, for the ALICE Collaboration} 

\affiliation{organization={Department of Physics},
            addressline={University of Jammu}, 
            city={Jammu},
            postcode={180006}, 
            state={Jammu and Kashmir},
            country={India}}

\begin{abstract}
Heavy quarks, such as charm and beauty, possess masses significantly larger than the characteristic energy scale of Quantum Chromodynamics (QCD), and are thus predominantly produced in hard-scattering processes with large momentum transfer ($Q^{2}$). This makes them effective probes for verifying QCD calculations and investigating heavy-quark hadronization processes across various collision systems. Recent measurements of charm meson and baryon production in proton--proton (pp) collisions at the LHC have enabled, for the first time, a determination of the total charm production cross section, including contributions from all ground-state charm hadrons. Additionally, the fragmentation fractions of charm quarks into hadrons in pp collisions have been quantified. Notably, the observed charm baryon-to-meson yield ratios are significantly enhanced compared to those measured in electron–positron and electron–proton collisions, indicating that charm quark hadronization is not universal across different collision systems. These findings highlight the importance of precise measurements of charm hadron production for understanding the hadronization mechanism in a parton-rich environment. This contribution presents the latest results on charm meson and baryon production in pp collisions, along with comparisons to theoretical models aimed at improving our understanding of charm quark hadronization.
  
\end{abstract}


\end{frontmatter}



\section{Introduction}
\label{sec1}
In proton–proton (pp) collisions at the LHC, heavy quarks—charm and beauty—are produced predominantly in the initial stages of the interaction through hard partonic scatterings, a process that can be accurately described by perturbative Quantum Chromodynamics (pQCD). The corresponding production cross section is calculated using a QCD factorization approach, which involves the convolution of three key components~\cite{1}:
(i) the parton distribution functions (PDFs), which represent the probability of finding a parton (quark or gluon) within the proton carrying a specific fraction of its momentum;
(ii) the hard-scattering cross section at the parton level, governing the dynamics of the heavy-quark production; and
(iii) the fragmentation functions, which encode the probability that a heavy quark transforms into a particular heavy-flavor hadron—such as a meson or baryon—carrying a defined momentum fraction.

As the PDFs and hard-scattering cross sections are common to all charm hadron species for a given collision energy and system, a common approach to studying charm-quark hadronization involves measuring the production yield ratios of different hadron species. In these ratios, the influence of the PDFs and partonic cross sections  cancels, making them sensitive probes of hadronization dynamics.
In this contribution, the production measurement of various charm mesons and charm baryons, using the data from pp collisions at $\sqrt{s}$ = 5.02 and 13 TeV, collected by the ALICE experiment during the LHC Run 2 are presented.


\section{Measurement of charm-hadron production with ALICE}
\subsection{Meson-to-meson ratio}
\begin{figure}[H]
   \begin{subfigure}[h]{0.5\textwidth}
 \includegraphics[width=\textwidth]{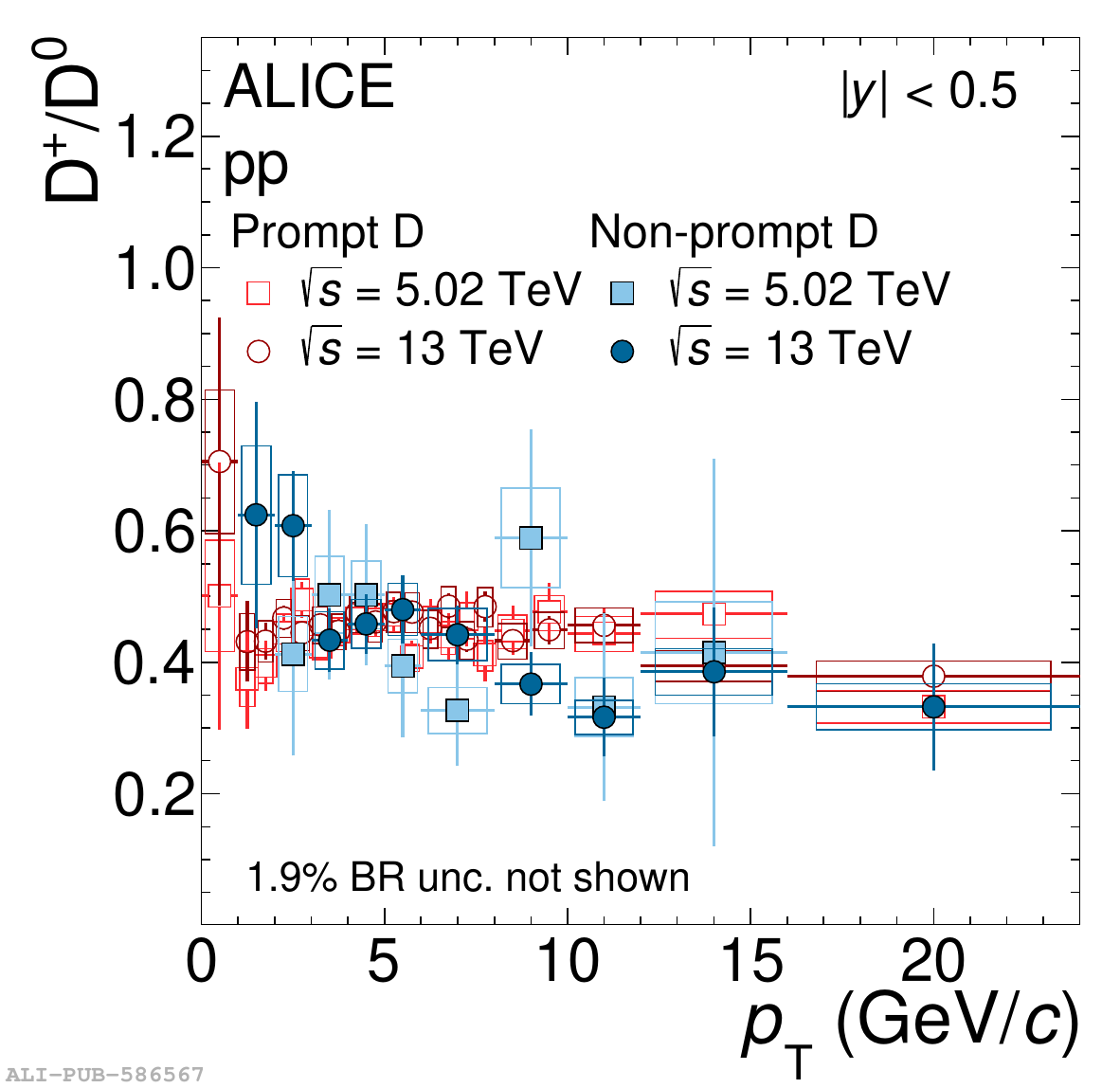}
   \end{subfigure} \begin{subfigure}[h]{0.5\textwidth}

\includegraphics[width=\textwidth]{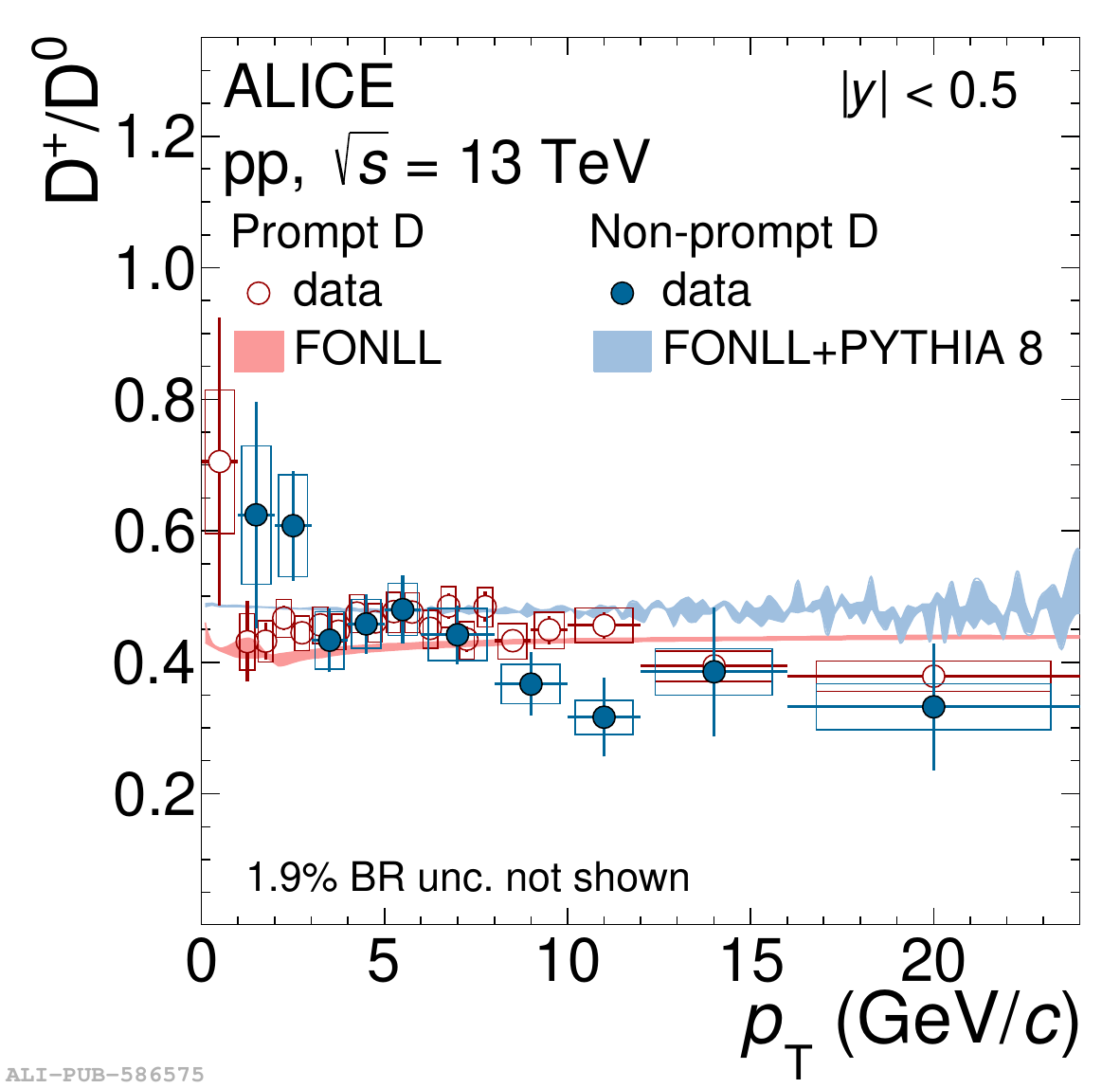}
   \end{subfigure}

\caption{Left: Prompt and non-prompt $\rm D^{+}$/$\rm D^{0}$ yield ratio as a function of $p_{\rm T}$ at $\sqrt{s}$ = 5.02 and 13 TeV. Right: Prompt and non-prompt $\rm D^{+}$/$\rm D^{0}$ yield ratio measured at $\sqrt{s}$ = 5.02 and 13 TeV compared to Fixed-Order plus Next-to-Leading Logarithmic (FONLL) calculations~\cite{2}.}\label{fig1}
\end{figure}

Figure~\ref{fig1} presents the ratios of prompt and non-prompt D-meson production cross sections measured in pp collisions at $\sqrt{s}$ = 5.02 and 13 TeV~\cite{2,4}. These results are compared with theoretical calculations based on pQCD: FONLL~\cite{1} for prompt D mesons and a combination of FONLL with PYTHIA 8~\cite{3} for non-prompt D mesons.  The ratios are consistent for both prompt and non-prompt D mesons, and the data are well described by FONLL calculations, which rely on quark fragmentation functions extracted from $\rm e^{+} e^{-}$ and ep collisions. This supports the assumption of universality of fragmentation functions for mesons. The results are also compatible at different center-of-mass energies.

\subsection{Baryon-to-meson ratio}
\begin{figure}[H]
   \begin{subfigure}[h]{0.5\textwidth}
 \includegraphics[width=\textwidth]{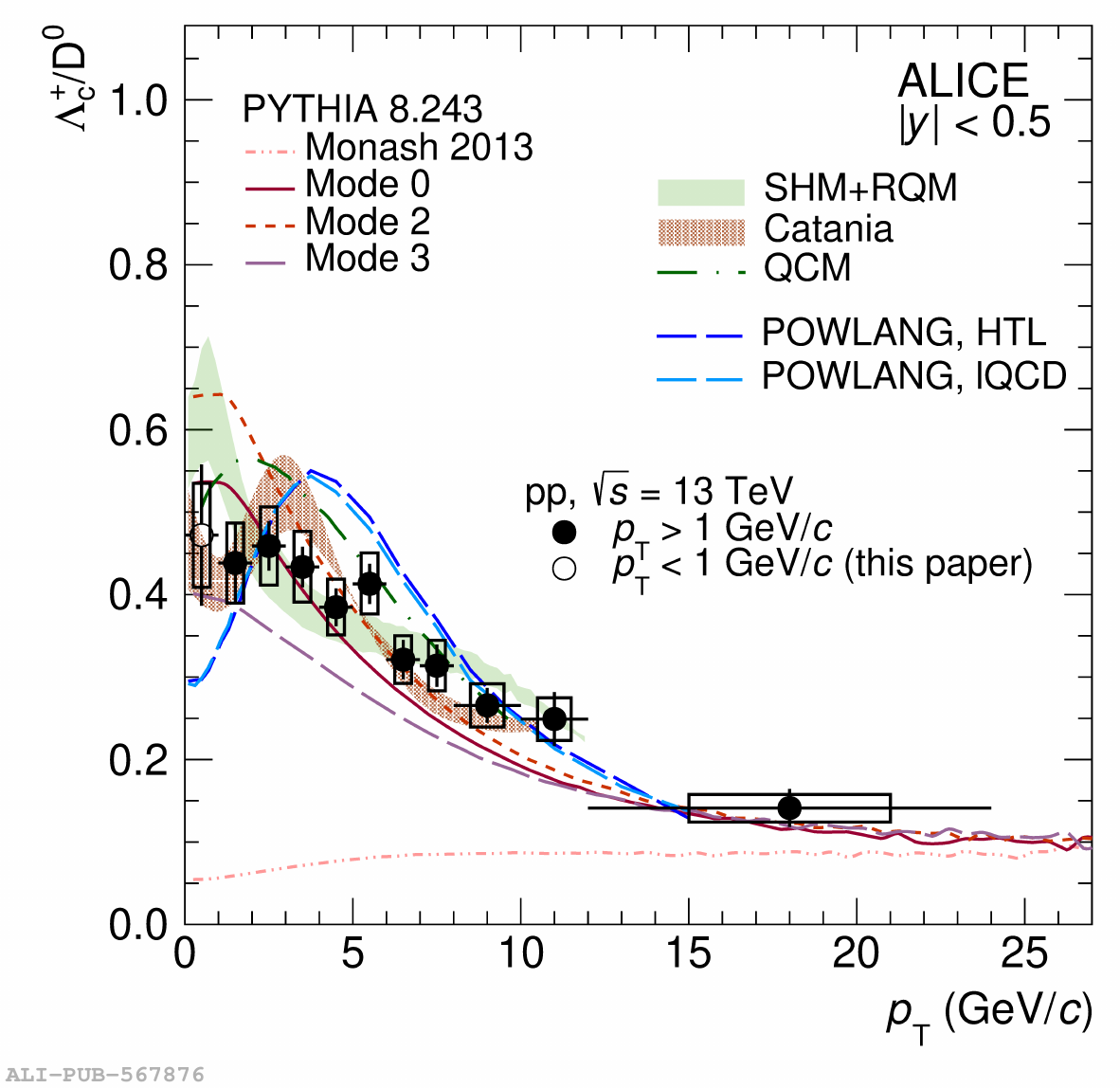}
   \end{subfigure} \begin{subfigure}[h]{0.5\textwidth}

\includegraphics[width=\textwidth]{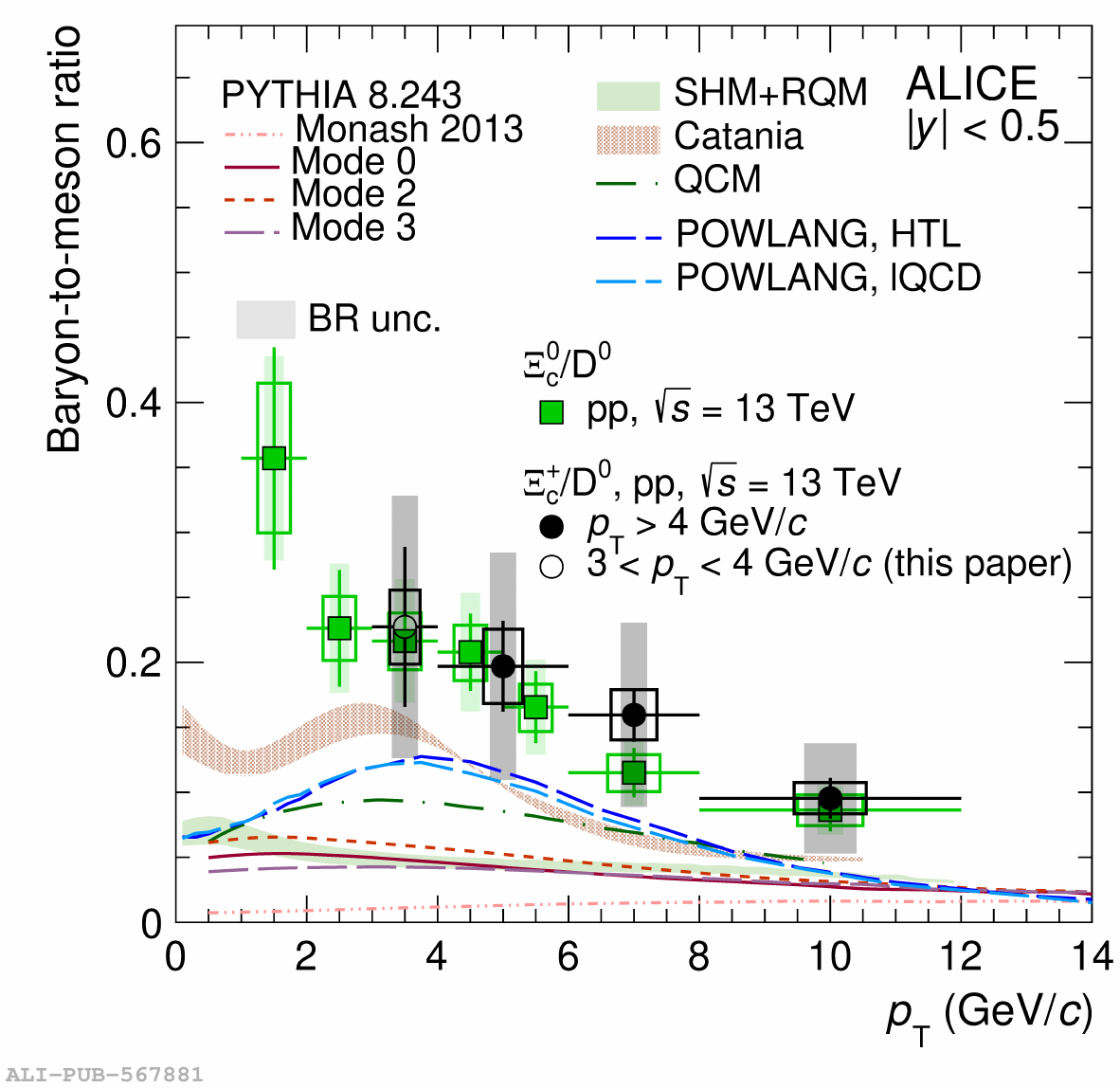}
   \end{subfigure}
\caption{Left: $p_{\rm T}$-differential prompt $\rm \Lambda_{c}^{+}$/$\rm D^{0}$ production yield ratio in pp collisions at $\sqrt{s}$ = 13 TeV compared with calculations from models. Right: $p_{\rm T}$-differential $\rm \Xi_{c}^{0}$/$\rm D^{0}$ and  $\rm \Xi_{c}^{+}$/$\rm D^{0}$ yield ratios in pp collisions at $\sqrt{s}$ = 13 TeV compared with model calculations ~\cite{6}.}\label{fig2}
   
\end{figure}

In contrast to the meson-to-meson ratios, the prompt $\rm \Lambda_{c}^{+}$/$\rm D^{0}$ yield ratio as a function of transverse momentum ($p_{\rm T}$), shown in the left panel of Figure~\ref{fig2}, reveals a pronounced $p_{\rm T}$ dependence ~\cite{5, 6}. The existing theoretical models significantly underestimate the charm baryon-to-meson yield ratios observed in hadronic collisions. A substantial enhancement at low and intermediate $p_{\rm T}$ is observed relative to model expectations based on fragmentation functions extracted from $\rm e^{+} e^{-}$ collisions, such as PYTHIA 8 with the Monash tune.

The enhancement of charm baryon production relative to mesons, as observed in hadronic collisions, has motivated the development of several theoretical models aimed at improving the description of the hadronization process. These include:

{\bf Color reconnection in PYTHIA 8}, incorporating color topologies with string junctions that enhance baryon production beyond the leading-color approximation~\cite{7};

{\bf The Statistical Hadronization Model (SHM)} coupled with the Relativistic Quark Model (RQM) ~\cite{8}, which includes feed-down contributions to $\rm \Lambda_{c}^{+}$ from an expanded spectrum of higher-mass charm baryons predicted by the RQM;

{\bf The Catania model~\cite{9} and the Quark Combination Model (QCM) ~\cite{10}}, both of which assume hadronization via quark coalescence, resulting in an enhanced baryon yield.

An enhancement relative to $\rm e^{+} e^{-}$ collisions is also evident in the $\rm \Xi_{c}^{0,+}$/$\rm D^{0}$ ~\cite{6, 11} yield ratios, as shown in the right panel of Figure \ref{fig2}. However, the discrepancies between experimental data and model predictions are more pronounced for charm-strange baryons than for non-strange baryons. This suggests that additional mechanisms enhance their production yields compared to those of non-strange charm baryons.  The Catania~\cite{9} and POWLANG~\cite{12} models are closer to the measured $\rm \Xi_{c}^{0,+}$/$\rm D^{0}$ ratio.

\subsection{Charm fragmentation fraction and total production cross section}



\begin{figure}[H]
   \begin{subfigure}[h]{0.5\textwidth}
 \includegraphics[width=\textwidth]{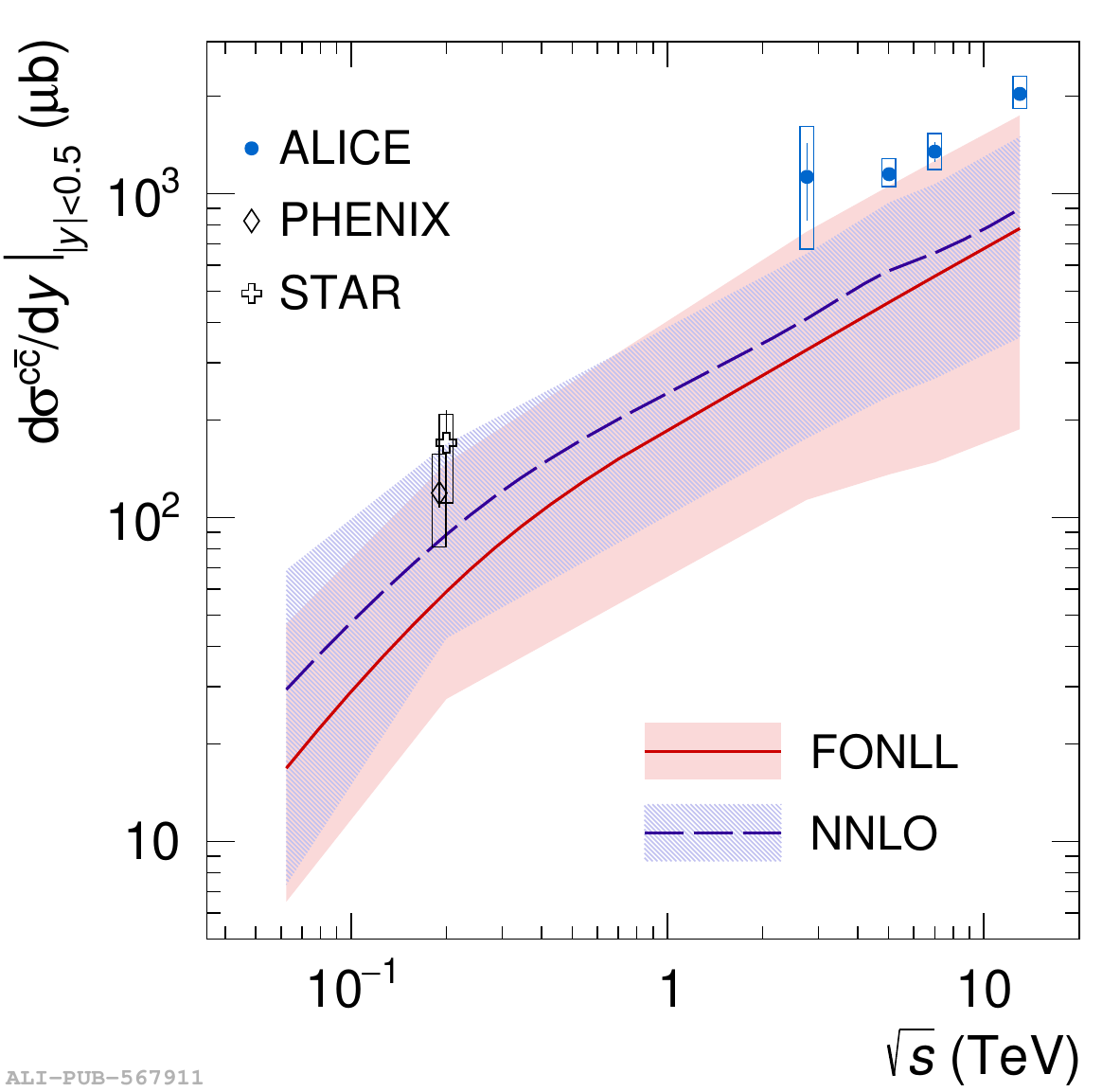}
   \end{subfigure} \begin{subfigure}[h]{0.5\textwidth}

\includegraphics[width=\textwidth]{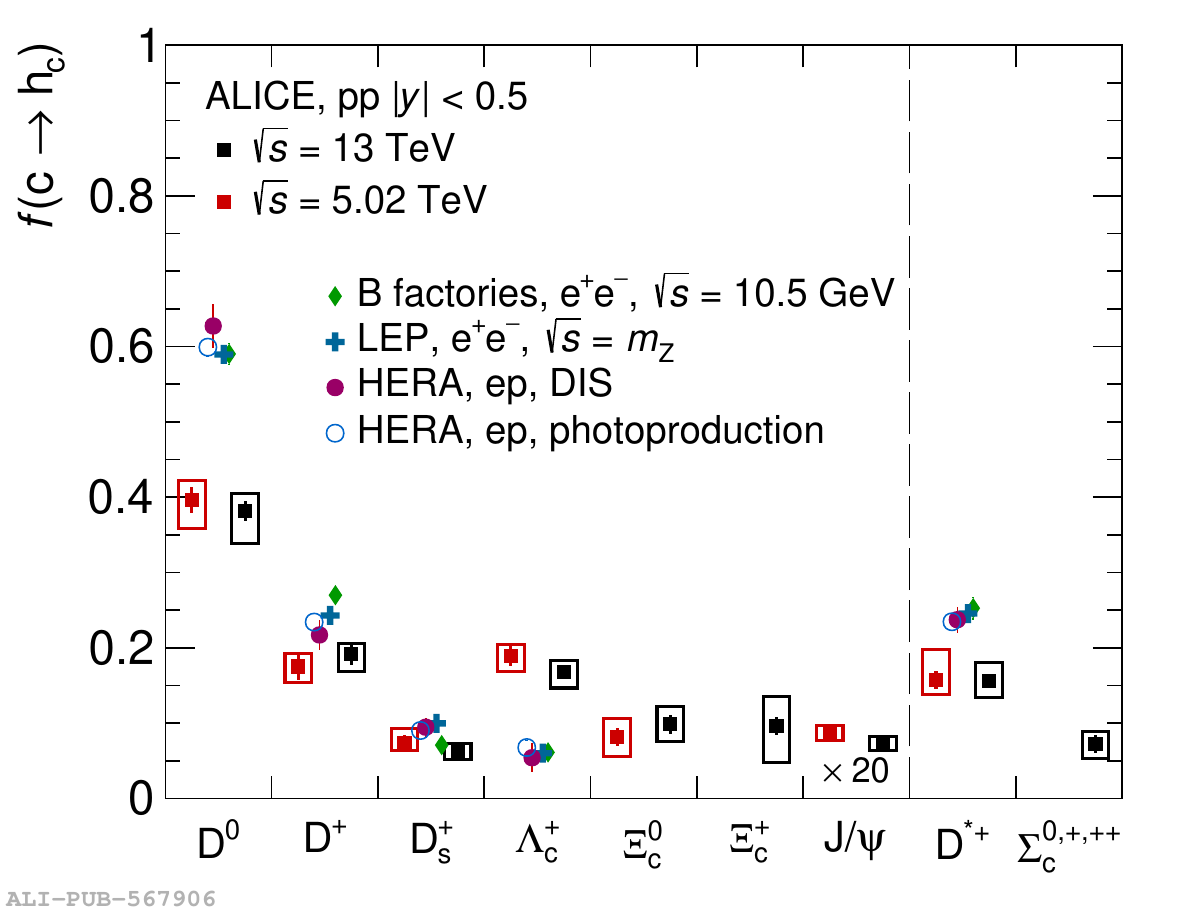}
   \end{subfigure}
\caption{Left: Total c$\rm \bar{c}$  production cross section at midrapidity per unit of rapidity as a function of the collision energy compared to FONLL~\cite{13} and NNLO~\cite{14}
calculations. Right: Charm-quark fragmentation fractions at midrapidity ($|y| <$ 0.5) in pp collisions at $\sqrt{s}$ = 5.02 and 13 TeV~\cite{6}.}\label{fig3}
   
\end{figure}

The right panel of Figure~\ref{fig3} presents the fragmentation fractions of charm quarks into various hadron species in pp collisions at $\sqrt{s}$ = 5.02 and 13 TeV. These results show no significant energy dependence within current experimental uncertainties. However, a clear difference is observed compared to fragmentation fractions extracted from  $\rm e^{+} e^{-}$ collisions, highlighting the role of the collision environment in charm-quark hadronization.

 Figure~\ref{fig3} (left) shows the total c$\rm \bar{c}$ production cross section at midrapidity ($|y| <$ 0.5), obtained as the sum of the production cross sections at midrapidity of the measured ground-state charm hadrons in pp collisions, at various center-of-mass energies. The ALICE measurements are compatible with FONLL calculations within 1--1.4 $\rm \sigma$ depending on the center-of-mass energy.

\section{Summary and outlook}
Heavy-quark production in pp collisions at the LHC provides key insights into QCD dynamics. ALICE has conducted extensive charm-baryon measurements, revealing intriguing findings. While meson-to-meson ratios exhibit no significant deviation from results obtained in $\rm e^{+} e^{-}$ and ep collisions supporting the  assumption of universality of the fragmentation functions, the baryon-to-meson ratios display a pronounced $p_{\rm T}$ dependence and a marked enhancement of baryon production, indicating a breakdown of universality. The ongoing LHC Run 3, with its significantly larger data sample and improved precision enabled by the upgraded detector systems, is expected to deliver new insights and refine the measurements achieved during Run 2.


\begin{thebibliography}{00}

\bibitem{1} M. Cacciari et al, JHEP 05 (1998) 007
\bibitem{2} ALICE Collaboration, JHEP 05 (2021) 220
  
\bibitem{3} M. Cacciari et al, JHEP 03 (2001) 006

\bibitem{4} ALICE Collaboration, JHEP 10 (2024)
\bibitem{5} ALICE Collaboration, Phys. Rev. Lett. 128 (2022) 012001
  \bibitem{6} ALICE Collaboration, JHEP 12 (2023) 086
\bibitem{7} J. R. Christiansen and P. Z. Skands, JHEP 08, 003 (2015)

\bibitem{8} M. He and R. Rapp, Phys. Lett. B 795, 117–121 (2019).
  
\bibitem{9} V. Minissale, S. Plumari, and V. Greco, Phys. Lett. B 821, 136622 (2021).
\bibitem{10} H.-h. Li, F.-l. Shao, and J. Song, Chin. Phys. C 45, 113105 (2021).
\bibitem{11} A. Beraudo et. al, Phys. Rev. D 109, L011501 (2024).
 \bibitem{12} ALICE Collaboration,  Phys. Rev. Lett. 127 (2021) 272001


  \bibitem{13} M. Cacciari et al., JHEP 10, 137 (2012)
 \bibitem{14} D. d’Enterria, A.M. Snigirev, Phys. Rev. Lett. 118, 122001 (2017)




  


\end{thebibliography}



\end{document}